\documentclass{PoS}

\title{Recent Developments in Dual Lattice Algorithms}

\ShortTitle{Dual Lattice Algorithms}

\author{\speaker{J. Wade Cherrington}\\ 
        University of Western Ontario\\
        E-mail: \email{jcherrin@uwo.ca}}

\abstract{We review recent progress in numerical simulations with dually transformed $SU(2)$ LGT,
starting with a discussion of explicit dual amplitudes and algorithms for $SU(2)$ pure 
Yang Mills in $D=3$ and $D=4$. In the $D=3$ case, we discuss results that validate the dual
algorithm against conventional simulations. We also review how a local, exact dynamical fermion 
algorithm can naturally be incorporated into the dual framework. We conclude with an outlook for 
this technique and a look at some of the current challenges we've encountered with this
method, specifically critical slowing down and the sign problem.}

\FullConference{The XXVI International Symposium on Lattice Field Theory\\
		 July 14-19 2008\\
		 Williamsburg, Virginia, USA}

\begin{document}

\section{Introduction: Duality on the lattice}
The general notion of duality in lattice gauge theory has been present in some form since the inception 
of the field\footnote{Early forerunners to the dual models discussed below were the strong coupling expansions 
used for example in~\cite{DrouffeZuber} and discussed further in~\cite{MandM}
and references therein.}.  Abelian dualities possess several generic features~\cite{Savit80},  including
the interchange of high and low temperature regimes between the original model and its dual, and a 
new gauge symmetry possessed by the dual model closely related to the original gauge symmetry
(sometimes referred to as "self-duality"). While the high-low temperature duality is also present in non-abelian
dual models~\cite{OecklPfeiffer}, a gauge symmetry of the dual is not generally present in the non-abelian case.
Consequently, the known techniques for simulating a gauge field are not directly applicable to the dual theory, and thus novel
techniques are required in the non-abelian case. For a few recent examples of simulations in the dual $U(1)$ case,
the reader is referred to~\cite{Jersak,Panero2004,Panero2005,PollyWiese,Zack98PhysRev}. 

As we shall see below, there is a framework~\cite{OecklDGT, OecklPfeiffer} within which the formal 
construction of non-abelian duals is straightforward.
However,  at the level of practical computations one is still faced with both non-trivial amplitudes (particularly in
the $D=4$ case) and constraints on allowed configurations that can make it difficult to find ergodic moves. Thus, as an alternative means of 
simulation, progress in the non-abelian case has been much slower and to date the ultimate feasibility of this approach remains an
open question.  Notwithstanding, we believe the results reviewed here (some of them quite recent) indicate definite progress towards
evaluating the potential for dual non-abelian simulations.

In the following we review some of the history of the subject and briefly describe some recent advances in computational 
methods for non-abelian dual simulations. We begin by describing the
form of the duality in the non-abelian case for groups $SU(N)$ in any dimension, and a systematic 
method for defining the dual amplitude in terms of representation-theoretic quantities. 
For the group $SU(2)$, these methods have recently led to an ergodic algorithm and the derivation of explicit 
amplitudes in both $D=3$ and $D=4$. Further, the dual degrees of freedom for the pure gauge field (closed branched surfaces) 
couple to the dual polymer picture of fermions in a geometrically natural way in the non-abelian case.
On a cautionary note, there are features of the resulting dual models that present a challenge for
practical computations: the dual amplitude is in general not positive definite, and extending existing methods
to weak coupling is complicated by long auto-correlation times. We shall return to these difficulties 
toward the end of the paper.

\section{Construction of the non-abelian duals}

One of the first explicit constructions of a dual non-abelian model on the lattice was reported in the early 1990's 
by Anishetty~\emph{et al.}\ for the $D=3$, $SU(2)$ case~\cite{AS,ACSM}. While no simulations
were reported in this work, the dual model was explicitly given and its local, discrete structure clearly exhibited; the 
model was in principle computable provided an ergodic algorithm could be found to sample the dual ensemble.  A further 
theoretical development came in 1995, when Halliday and Suranyi~\cite{Halliday95} described the construction of $D$-dimensional
non-abelian dual models on the lattice in which the dual degrees of freedom are local. A few years later, an important series of papers by
Hari Dass \emph{et al.}~\cite{Dass83,Dass94,DassShin} described the first reported attempts at a full lattice simulation using the dual
amplitude presented by Anishetty~\emph{et al.} in~\cite{AS,ACSM}.

Further theoretical developments came with the work of Oeckl and Pfeiffer~\cite{OecklPfeiffer}. Here, a general 
non-abelian gauge theory is considered and the dual theory is constructed in terms of representation-theoretic
quantities. The degrees of freedom of the resulting dual theory are manifestly local, with a local amplitude defined in
terms of contractions of invariant tensors that define face, edge, and vertex amplitudes. Borrowing some terminology
from loop quantum gravity, this structure is sometimes referred to as a lattice \emph{spin foam} model. The reader is also referred 
to~\cite{OecklDGT} for a clear pedagogical treatment and a generalization of the framework that includes $q$-deformed and
supersymmetric models. 

For concreteness, we shall review here the explicit form of the duality. Recall that given an oriented lattice, the action is 
discretized by forming a sum over plaquettes, $S[g] = \sum_{p\in P} S(g_p)$, where the group element $g_p$ is the holonomy
around an oriented plaquette $p$, obtained by forming a suitable product of the variables $g_{e}$ and their inverses. 
The lattice partition function is
\begin{equation}\label{eq:conventional}
	\mathcal{Z} = \int \prod_{e\in E} dg_e\, e^{-\sum_{p\in P} S(g_p)},
\end{equation}
where $E$ is the set of lattice edges. To pass to the dual model, the amplitude at each plaquette is expanded into the 
characters of $G$:
\begin{equation}
e^{S(g_{p})}= \sum_{i} c_{i} \chi_{i} (g_{p}),
\label{character_expansion}
\end{equation}
where the sum is over all unitary irreducible representations (irreps) of the group $G$; $\chi_{i}$ is the character function on $G$
associated with the $i$th irrep. The next step in moving to a dual model is to interchange the order of integration and summation:
\begin{equation}
\mathcal{Z} = \int \left( \prod_{e\in E} dg_{e} \right) \prod_{p \in P} \sum_{i} c_{i} \chi_{i}(g_{p})
= \sum_{s:P \rightarrow \mathcal{J} } \int \left( \prod_{e \in E} dg_{e} \right) \prod_{p \in P} c_{s(p)} \chi_{s(p)}(g_{p}),
\label{interchange}
\end{equation}
where the maps $P \rightarrow \mathcal{J}$ are labellings of the plaquettes of the lattice by unitary irreducible 
representations of $G$. At this stage, the model can already be viewed as dual (the integrals being absorbed 
into the definition of the amplitude). However, this is of limited use computationally as the 
integrals are non-local, high dimensional integrals with no obvious closed form. 
The key insight made explicit by~\cite{OecklPfeiffer} is that the integrals
project onto a basis of invariant tensors (also referred to as~\emph{intertwiners}). Due to space limitations, we 
won't derive the rest of the duality transformation. The resulting dual model has the following form:
\begin{eqnarray}
\mathcal{Z}  = \sum_{s:P \rightarrow \mathcal{J} } \int \left( \prod_{e \in E} dg_{e} \right) \prod_{p \in P} c_{s(p)} \chi_{s(p)}(g_{p}) 
 = \sum_{f\in \mathcal{F}}\prod_{p \in P}A_{P}(f,p)\prod_{ e \in E}A_{E}(f,e)\prod_{v \in V}A_{V}(f,v)
\label{eq:ZVac}
\end{eqnarray}
where (for $D=3$), the amplitudes can be defined as:
\begin{equation}
A_{V}(f,v) = \raisebox{-1.40cm}{\includegraphics[scale=0.5]{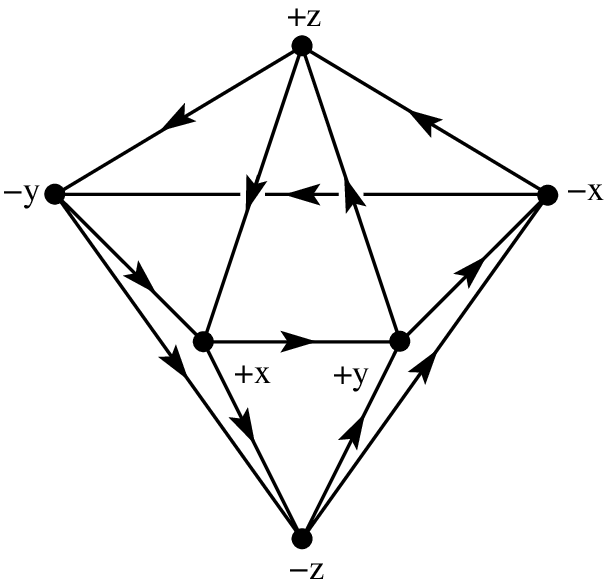}},  \hspace{0.6cm} A_{E}(f,e) = \raisebox{-.60cm}{\includegraphics[scale=0.8]{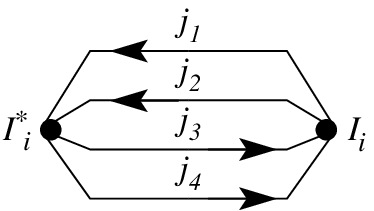}}, \hspace{0.5cm} A_{P}(f,j_{p})= c_{p},
\end{equation}
and $\mathcal{F}$ is the set of all labellings of the plaquettes by irreps of $G$ and the lattice edges by
intertwiners (for any plaquette labelling, these have a discrete basis indexed by a finite number of irreps). 
The arguments to the $A_{V}$ function are as follows: irrep labelling plaquettes incident to $v$ are assigned to edges, and intertwiners associated
to incident edges are assigned to vertices of the $A_{V}$ network.
In the case of $G=SU(2)$, a powerful diagrammatic calculus exists~\cite{KauffmanLins} for the explicit evaluation of quantities such as
$A_{V}(f,v)$ and $A_{E}(f,e)$.
It can further be shown that the non-zero contributions to the partition function have the geometric interpretation of closed,
branched surfaces colored by the irreps of $G$.

We mention in passing that while an explicit formula for the $D=3$ case has been known since~\cite{AS,ACSM}, an explicit form
for the $D=4$ vertex amplitude practical for simulations was not known until very recently with~\cite{CD4}. This amplitude
 (a function of 48 half-integer spin labels)
takes order $j^4$ time and makes it possible to investigate the possibility of dual simulations the four-dimensional case, ultimately the most
significant for physical models.  
\section{An ergodic non-abelian dual algorithm $SU(2)$ case}
\begin{figure}[h]
\begin{center}
\includegraphics[scale=0.78]{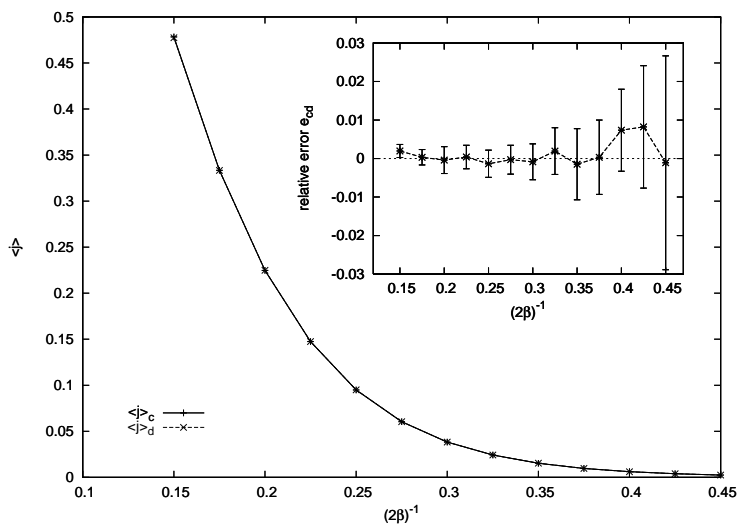}\hspace{0.7cm}
\end{center}
\vspace{-0.3cm}
\caption{Conventional and dual results for the $8^{3}$ lattice. }
\vspace{-0.1cm}
\label{fig:L8_results}
\end{figure}

In~\cite{CCK} the author and collaborators constructed an ergodic algorithm for sampling the dual configurations. In the case $G=SU(2)$ and
$D=3$, we were able to obtain high quality data up to a $\beta$ of 2.85.  The results of our runs (using $8^3$ and $16^3$ lattices)
were in statistical agreement with data obtained by conventional means, which we used as a check on our calculations.
The expectation value of spin (averaged over the lattice) was used as an observable; a plot of our simulations including residuals is shown in 
Figure~\ref{fig:L8_results}.
While~\cite{CCK} focused on the $D=3$, $SU(2)$ case, the algorithm developed has a straightforward generalization with respect to group and dimension.
Currently simulations in $D=4$ are being tested.

While our first tests used spin as an observable, it is straightforward to compute Wilson loop
observables in the dual; in fact, a method very much analogous to a dual $U(1)$ used by Panero~\cite{Panero2005} can be applied to build up the 
expectation values of Wilson loop observables. In the abelian case, this method eliminates exponential decrease in signal to noise,
and should behave the same in the non-abelian case 
\section{Dynamical fermions in the dual}
Recall that when coupling gauge fields to fermions on the lattice, an exact integration of the fermionic Grassman variables yields
an effective action for the gauge fields given by the determinant of the quark matrix. This determinant is a non-local function of
the gauge degrees of freedom and consequently a considerable challenge to compute with~\cite{DeDe}.
This determinant can in turn be expanded into contributions associated to closed polymers on the lattice.
The polymer expansion can be viewed as a duality for pure lattice fermions on the lattice, and indeed has been pursued in that
context as an alternative computational framework~\cite{KST,Montvay90}. The usefulness of this method is best established in two
dimensions (see, for example~\cite{FortSims, Wolff}; in three and four dimensions, the presence of an alternating sign in the
amplitude presents a challenge). 

\begin{figure}[h]
\begin{center}
\includegraphics[scale=0.6]{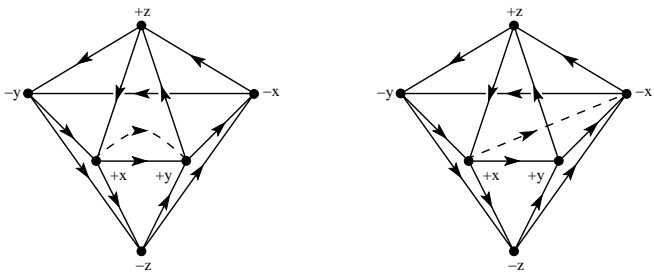} 
\end{center}
\vspace{-0.4cm}
\caption{Charged vertex amplitudes arising in dual dynamical fermion model}
\vspace{-0.4cm}
\label{fig:fermions}
\end{figure}

One can show~\cite{fermions} that applying the spin foam expansion of the gauge degrees of freedom (\ref{eq:ZVac}) to each term 
in the polymer expansion results in a configuration space of closed branched surfaces ending on the one-dimensional closed polymers.
Remarkably\footnote{Note however an analogous construction for the $U(1)$ case had been made previously in~\cite{Fort}.}, the resulting 
amplitude that combines factors from the spin foam and polymer expansion is local in the spin foam and fermion degrees
of freedom.  The factoring of the amplitude into plaquette, edge, and vertex amplitudes remains unchanged, the main difference being 
that the vertex amplitudes at vertices along the fermionic loop are modified by the presence of charge running through that vertex.  
As in the pure gauge theory, the modified vertex amplitude can be defined in terms of a spin network evaluation; examples of modified 
vertex amplitudes are shown in Figure~\ref{fig:fermions}.

\section{Challenges and Outlook}
\emph{The sign problem} --- If the expectation value of the sign of the amplitude
is close to zero then methods for extracting expectation values (i.e. "sign trick", see~\cite{Dass83}) 
which rely on dividing by the sign of expectation value will exhibit large numerical errors.
In the simulations performed in~\cite{CCK}, the presence of a negative amplitude had to be accounted for, however the expectation
value of the sign remained close to unity, decreasing gradually towards weaker coupling.  As weaker coupling values were used, long auto-correlation
times (discussed next) made it difficult to assess the effect of mixed signs in the amplitude.  The 
immediate question under investigation is what the status of the sign problem is in four dimensions, as 
a function of coupling. Depending on whether the naive "sign trick" method is successful at couplings used 
for practical simulations, new methods may need to be developed.

\emph{Auto-correlation at weak coupling} --- As one approaches weaker coupling, the dominant contributions transition from the isolated bubbles 
characteristic of strong coupling, to a highly disordered phase of surfaces extending throughout the lattice.  This suggests that a cluster algorithm capable of evolving
state on larger scales may be a feasible strategy. Of particular interest in this connection is a recent proposal by Wolff~\cite{WolffWorm} to
apply a worm algorithm in a dual setting, which demonstrates very promising critical behavior for a $Z_2$ dual model.

\emph{Outlook} --- Current computational efforts of the author and his group are focused on optimizing the $D=4$ algorithm, addressing 
the sign problem in various dual models, and developing cluster methods to improve auto-correlation times at weak coupling. 
On the theoretical front, we are also developing methods for evaluating $G=SU(3)$ models, in which finding explicit amplitudes remains an
open problem.  In the near future, we will be investigating incorporating supersymmetry using dual methods, as well as dual 
approaches to chiral fermion simulations.

While we remain cautious in light of the remaining questions with regard to the sign problem, the dual approach continues to 
provide an intriguing alternative perspective on the physics of lattice gauge theory.

{\bf Acknowledgements} --- The author would like to thank Dan Christensen and Igor Khavkine for collaboration on
work dicussed here, as well as Florian Conrady and Marco Panero for valuable discussions 
relating to this topic. The author is supported by an NSERC postgraduate scholarship. This work was made possible 
by the facilities of the Shared Hierarchical Academic Research Computing Network (SHARCNET).

\end{document}